# Magnetically controlled vector based on E coli Nissle 1917

S.V. Gorobets[1], O.Yu. Gorobets[1,2*], I.V. Sharau[2], Yu.V. Milenko[1]

[1]*National Technical University of Ukraine «Igor Sikorsky Kyiv Polytechnic Institute», 37 Peremohy Ave., 03056 Kyiv, Ukraine*

[2]*Institute of Magnetism of NAS and MES of Ukraine, 36b Acad. Vernadskoho Blvd., 03142 Kyiv, Ukraine*

*Correspondent author e-mail: gorobets.oksana@gmail.com, National Technical University of Ukraine «Igor Sikorsky Kyiv Polytechnic Institute», 37 Peremohy Ave., 03056 Kyiv, Ukraine

## Introduction

For decades gene or protein replacement therapy has been proposed as means of preventing and treating various human diseases, in particular cancer [1] and monogenic diseases such as hemophilia and cystic disease, fibrosis [2–5]. The delivery of proteins or nucleic acids to the target cells is carried out through various mechanisms; these include viral vectors, electroporation, microinjection, lipofection, and others [6,7]. In the past, most researchers have focused on the use of viral vectors that have high delivery efficiency for gene therapy [6]. However, some drawbacks have been noted for viral transduction methods, including: primarily safety of use [8,9], high cost, short-lived bioactivity, size limitation for DNA payload, and problems with immunogenicity and cytotoxicity [6,10,11]. As a result, alternative methods of creating vectors for targeted drug and DNA delivery are being explored. In the mid-1990s, the benefits of using bacterial carriers as vectors for delivery of eukaryotic plasmids were introduced with different bacteria being investigated such as *Shigella*, *S. typhimurium*, *Salmonellatyphi*, *S. flexneri*, *L. Monocytogenes*, *E. coli* and others [12–17]. *E. coli* is an integral part of the human gastrointestinal flora and is therefore considered as an alternative for delivery through the gut when using gene therapy. The ability of any bacteria, including *E. coli*, to act as vectors for delivery includes the uptake of target cells and the release of therapeutic load from vacuoles to cytosol. Professional phagocytic cells, such as neutrophils, dendritic cells and macrophages, absorb bacterial vectors through phagocytosis. Non-professional phagocytes (such as epithelial cells), which are often involved in disease processes, may be promising for some therapies and designed to actively interact with these cells [5].

Unlike *S. typhimurium* and other bacteria that colonize tumors with necrosis, *E. Coli* Nissle 1917 (EcN) has a higher ability to target the tumor, as it mainly propagates in the area between necrotic and hypoxic regions of tumors [18–20], which guarantees the penetration into the hypoxic areas of the drug, which is loaded into EcN. An additional advantage of EcN is that the cell membrane of EcN can directly interact with the adaptive immune system and therefore reduces inflammation [21,22]. Moreover, serum sensitive lipopolysaccharide (LPS) of the EcN membrane provides rapid removal of this strain from normal organs and tissues. Therefore, EcN-like cells with a specific immune regulatory system and capable of targeting hypoxic sites have potential applications for transporting chemotherapeutic agents to the depth of the tumor site [23].

The newest direction in the construction of bacterial vectors is, first, artificial magnetic labeling of bacteria, and, second, the use of microorganisms with natural magnetically controlled properties in order to combine targeting under the influence of a chemoattractant gradient and an external magnetic field gradient [24]. Thus, an additional advantage of using magnetically driven bacteria for drug delivery is that the application of a magnetic field can be used to reach a specific target tissue in the body without affecting other non-target tissues. An example of studies in the first direction is the manufacture of stochastic "microcarriers" that move at an average speed of up to 22.5 μm/s by attaching drug-loaded microparticles to embedded magnetic nanoparticles [25]. These "microcarriers" exhibit directional motion under the gradient of the chemoattractant and the magnetic field, respectively. Thus, bacterial-controlled multifunctional magnetic biosensors can be used for targeted delivery of drugs with significantly enhanced efficacy compared to passive microparticles or bacterial vectors without magnetically controlled properties.

An interesting example of a "microcarrier" in the second direction is given in [26]. Magnetotactic bacteria *Magnetospirillum gryphiswalense* (MSR-1) loaded with antibiotic-coupled mesoporous silica microtubes (biocomposite) were targeted to the infectious biofilm. A complex of MSR-1 cells with attached antibiotic biocomposite particles was delivered to a mature *E. coli* biofilm, combining the ability of the MSR-1 cells to move and their magnetic properties with subsequent antibiotic release and destruction. As a result, the potential for the use of magnetically driven bacterial vectors has been identified. In addition, the use of whole cells of magnetotactic bacteria and magnetosomes as their constituent parts is discusses as "reasonable therapeutic agents" for an effective delivery system targeted at a specific site or organ in the body [27,28]. In [28], it was shown that the magneto-aerotactic migration behavior of magnetotactic bacteria, *Magnetococcus marinus* MC-1, can be used to transport drug-loaded nanoliposomes into hypoxic regions of the tumor characterized by low oxygen levels and, as a rule, resistant to anti-cancer therapy. Each MC-1 cell contains chains of magnetic iron-oxide nanocrystals in natural medium and typically move along magnetic field lines and toward low oxygen concentrations based on a dual system of magneto-aerotactic behavior. Approximately 70 drug loaded nanoliposomes were attached to each MC-1 cell c. Covalently bound cells of MC-1 nanoliposomes containing drugs were introduced near the tumor c. Magnetococcus marinus MC-1 cells were found to be alive and mobile and exhibited both magnetotaxis and aerotaxis responses after administration to mice in the peritomoral region, reaching deeper tumor areas compared to passive agents (microspheres and dead *Magnetococcus marinus* cells). Magnetic control resulted in up to 55% of MC-1 cells penetration into hypoxic areas in mice. These results suggest that it is possible to use microorganisms that exhibit magneto-aerotactic behavior, and thus significantly improve therapeutic performance in hypoxic areas of tumors. *Magnetococcus marinus* MC-1 is a marine bacterium that was not expected to survive in mammals, although this has not been investigated in [28]. The cells of *Magnetococcus marinus* strain MC-1 are clinically "safe", do not cause adverse effects when introduced into mice according to another result of the paper [28]. This result is unexpected due to the common immunogenic properties of the gram-negative bacterial cell wall [29,30]. Therefore, it is impossible to guarantee the safety of this method without further detailed studies on the effects of the introduction of magnetotactic bacteria into living organisms. This is probably the reason why, as a rule, not the magnetotactic bacteria, but the magnetosomes, as their constituents, are used as vectors for targeted drug delivery. Thus, these structures cannot reproduce, cannot cause infections, and do not produce a pronounced immunological response, since the magnetosome membrane does not have lipopolysaccharides on the outer membrane of the gram-negative cell wall, which are known to act as endotoxins [30,31]. However, the disadvantage of using magnetosomes as vectors is that they are incapable of chemotaxis, and their manufacturing technology is high-cost.

**Results and discussion**

This paper proposes a new magnetically guided bacterial vector based on EcN, which has all of the above benefits of using this particular *E. coli* strain, including aerotactic behavior, and which lacks the disadvantages of magnetotactic bacteria in terms of potential immunological response and complexity of cultivation in the laboratory. In this paper, it is first discovered that EcN has natural magnetically controlled properties, that is, EcN cells move in a gradient magnetic field of permanent magnets without artificial magnetic labeling. It is also shown in the paper that the cultivation of EcN in a medium enriched with iron chelates and under the influence of an external magnetic field increases several times the magnetophoretic mobility of EcN cells compared to cultivation under standard conditions. Thus, the speed of movement of EcN cells in a gradient magnetic field of laboratory magnets is achieved of the order of several mm/s. It has also been shown for the first time that the cultivation of EcN biomass is accelerated by cultivation in an external magnetic field, while the change in the concentration of iron chelates in the medium has no effect on the cultivation dynamics of the EcN culture.

The bioinformatics analysis using comparative genomics methods in the BLAST program of the US National Center for Biotechnology (NCBI) was performed to determine whether EcN has natural magnetically driven properties, i.e. whether EcN is a producer of biogenic magnetic nanoparticles (BMNs). The EcN proteome was aligned with the amino acid sequences required for the biomineralization of Mam proteins of the magnetotactic bacterium *Magnetospirillum gryphiswaldense* MSR-1, for which BMN biomineralization has been studied in detail at the genetic level.

The standard parameters were taken into account in the alignments to evaluate the homology of the amino acid sequences of the EcN proteins with the MSR-1 proteins [32]. *E*-number is an indicator of the statistical significance of the alignment; Ident (*I*) is the percentage of identical amino acid residues in pairwise alignment of a given protein sequence; Length is length of alignment and functions of the proteins under study. Alignment results show that EcN is a BMN producer (Table. 1.).

Table 1. Alignment of the proteins of the Mam-group of *M. gryphiswaldense* MSR-1 and the proteome of the bacterium EcN.

| The strain of the microorganism | *E*-number (*I*, %) | | | | |
|---|---|---|---|---|---|
| | Proteins of *Magnetospirillum gryphiswaldense* MSR-1 | | | | |
| | MamA | MamB | MamM | MamO | MamE |
| *Escherichia coli* Nissle 1917 | 0.001 (23,86) | 5e-37 (30,54) | 2e-29 (29,89) | 3e-09 (29,70) | 1e-35 (40,38) |

Experimental confirmation of the presence of BMN in EcN cells was performed using atomic force microscopy (AFM) and magnetic force microscopy (MSM) methods (Fig. 1).

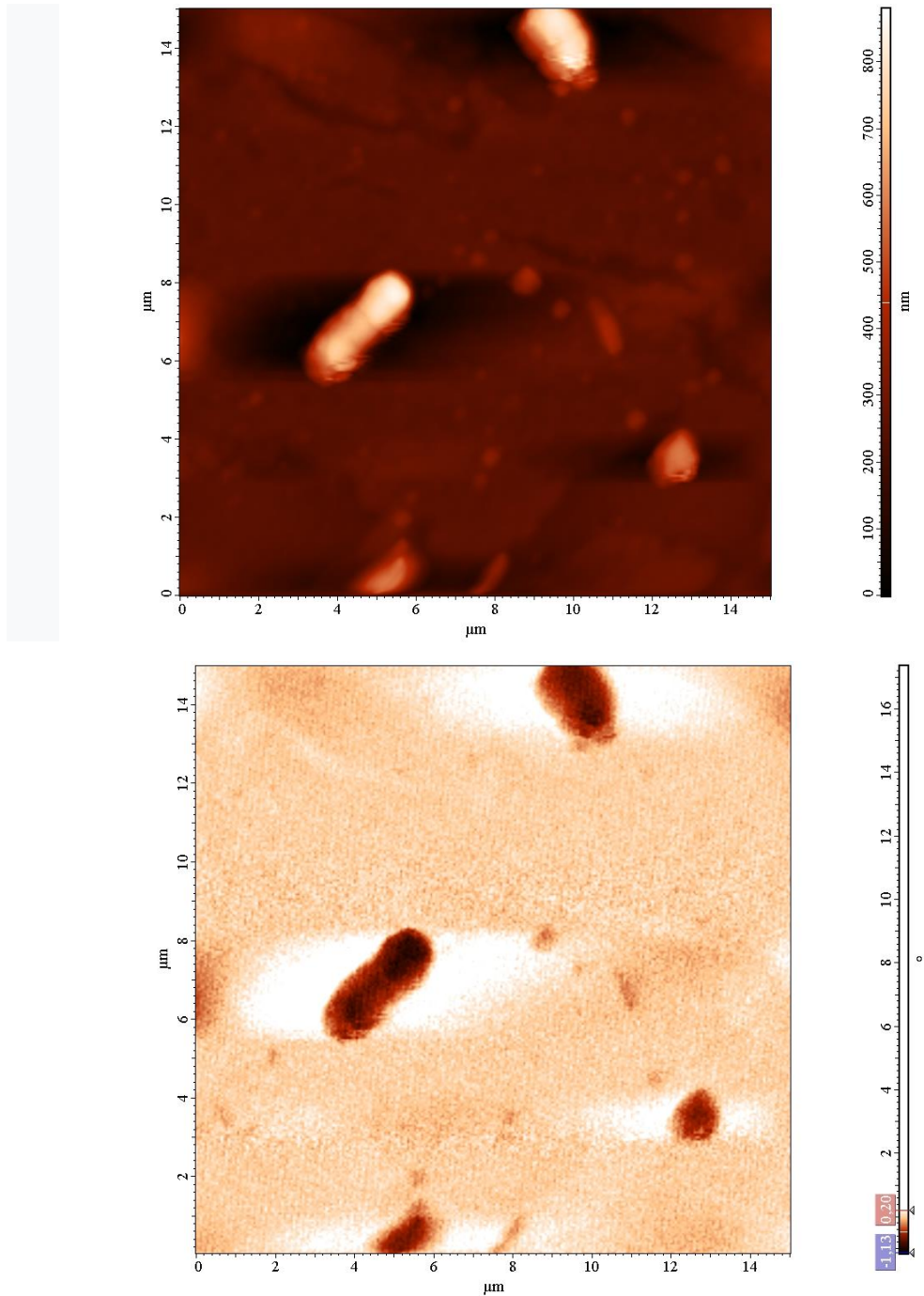

Fig. 1. AFM (up) i MFM (down) images of EcN cells cultivated on standard medium (control).

The MFM image of Fig. 1 shows that an uneven distribution of magnetic nanoparticles in the form of clusters of different sizes is observed inside the EcN cells, similar to the results of [33], where a study of BMN in Escherichia coli VKM B-126 was performed using transmission electron microscopy. This result, according to the results of bioinformatics analysis of this paper (Table 1) and the research [34] and experimental studies of this paper (Fig. 1) and the paper [33], indicates that BMN available in EcN cells are amorphous.

An aqueous solution of iron chelate (32 mg/ml) was added to intensify the process of BMN formation and to enhance the magnetically driven properties of EcN cells in standard EcN growing media such as meat-peptone agar (MFA) and meat-peptone broth (MFB). The EcN culture was cultivated on medium supplemented with iron chelate for two days under optimal temperature conditions of 37°C. After that, the morphological characteristics of BMN bacteria have been re-evaluated by means of the methods of AFM and MFM (Fig. 2).

As can be seen from Fig. 2, both the EcN cells and BMNs inside the EcN cells form chains after growing on a standard medium with the addition of iron chelate.

The bacterium EcN was cultivated under four different conditions for a detailed study of possible methods of changing of the magnetically controlled properties: a) standard medium, b) standard medium with the addition of an aqueous solution of iron chelate (32 mg/ml), c) standard medium under the influence of an external magnetic field (MF) with magnetic field flux density 0.15 T, and d) standard medium with the addition of an aqueous solution of iron chelate (32 mg/ml) under the influence of an external MF with a magnetic field flux density 0.15 T. Samples of suspensions of the four cultures under study were analyzed at the contact surface of the system of two NdFeB permanent magnets (Fig. 3).

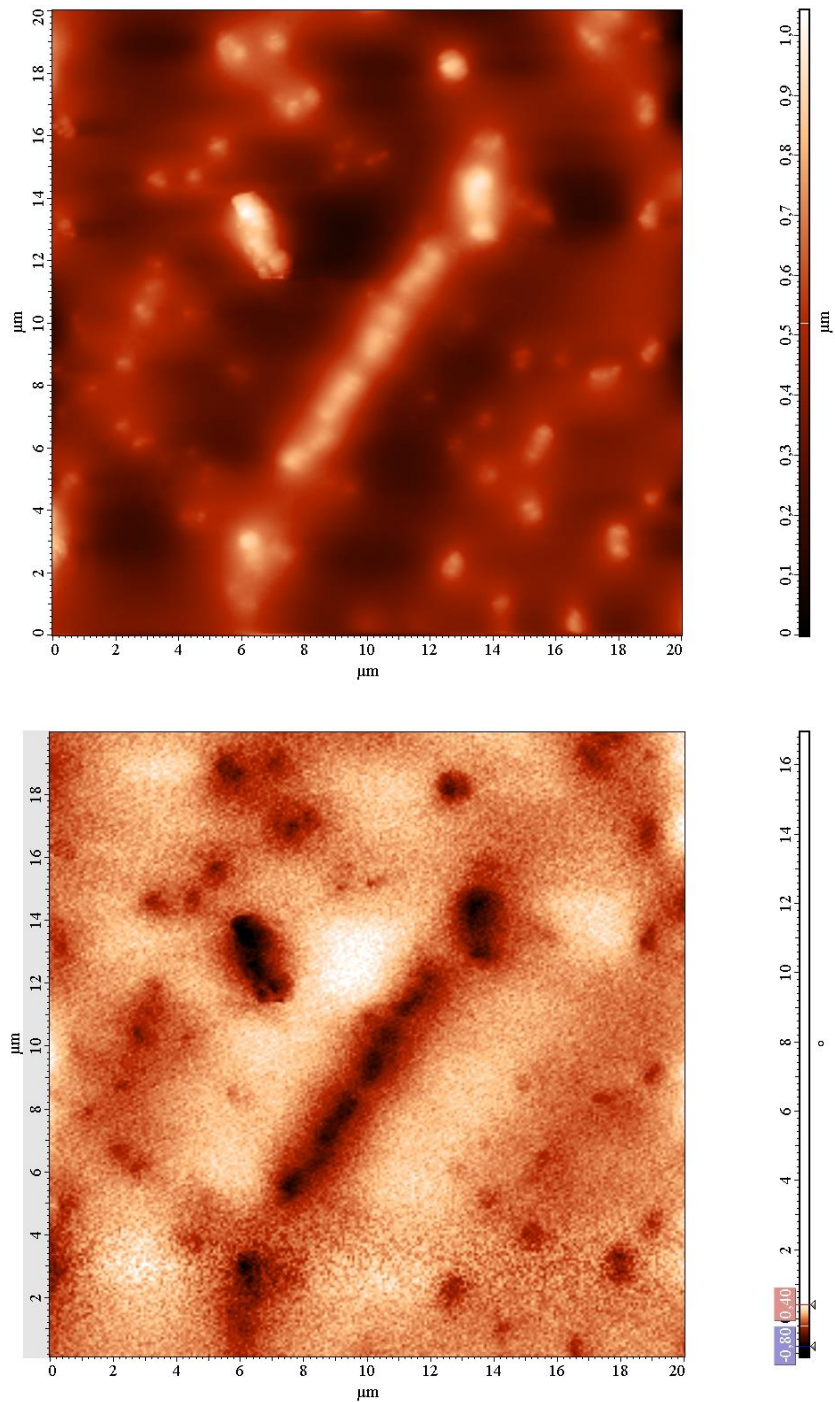

Fig. 2. AFM (up) and MSM (down) images of EcN cells cultivated on medium supplemented with iron chelate.

The EcN cells cultivated under the influence of an external MF manifest self-arrangement in chain-like structures (Fig. 2). Such arrangement is typical for objects possessing the remnant magnetization and interacting as magnetic dipoles.

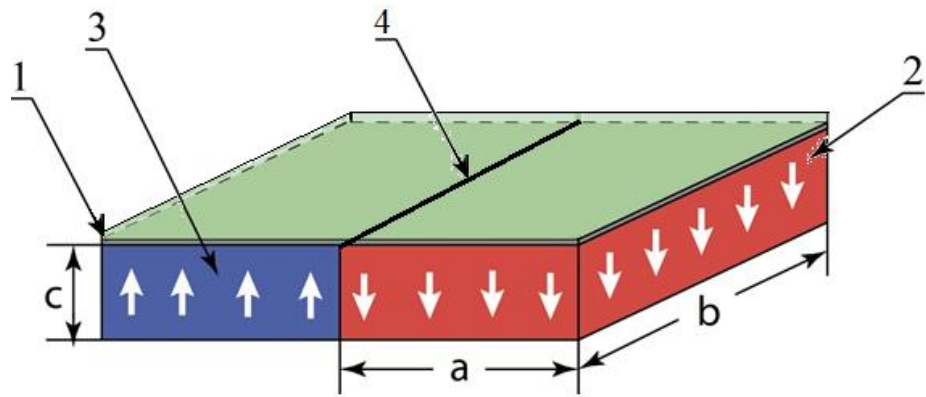

Fig. 3. The schematic image of the system of two NdFeB permanent magnets. 1 – cover glass; 2 – NdFeB permanent magnet with up orientation of magnetization as indicated by white arrows; 3 – NdFeB permanent magnet with down orientation of magnetization as indicated by white arrows; 4 – contact surface of two NdFeB permanent magnets; a=23 mm, b=30 mm, c=10 mm.

To do this, 0.02 ml of each suspension with a concentration of $3 \cdot 10^7$ cells/ml was applied to the cover glass (0.2 mm thickness), the glass was placed above the contact surface of the system of two permanent magnets (Fig. 3). The bandwidth (Fig. 4) formed in the inhomogeneous MF above the contact surface of the system of two permanent magnets varied significantly for each of the four studied cultures, which characterizes the magnetic interaction of BMN of bacteria with a high-gradient magnetic field (HGMF) created by the system of magnets.

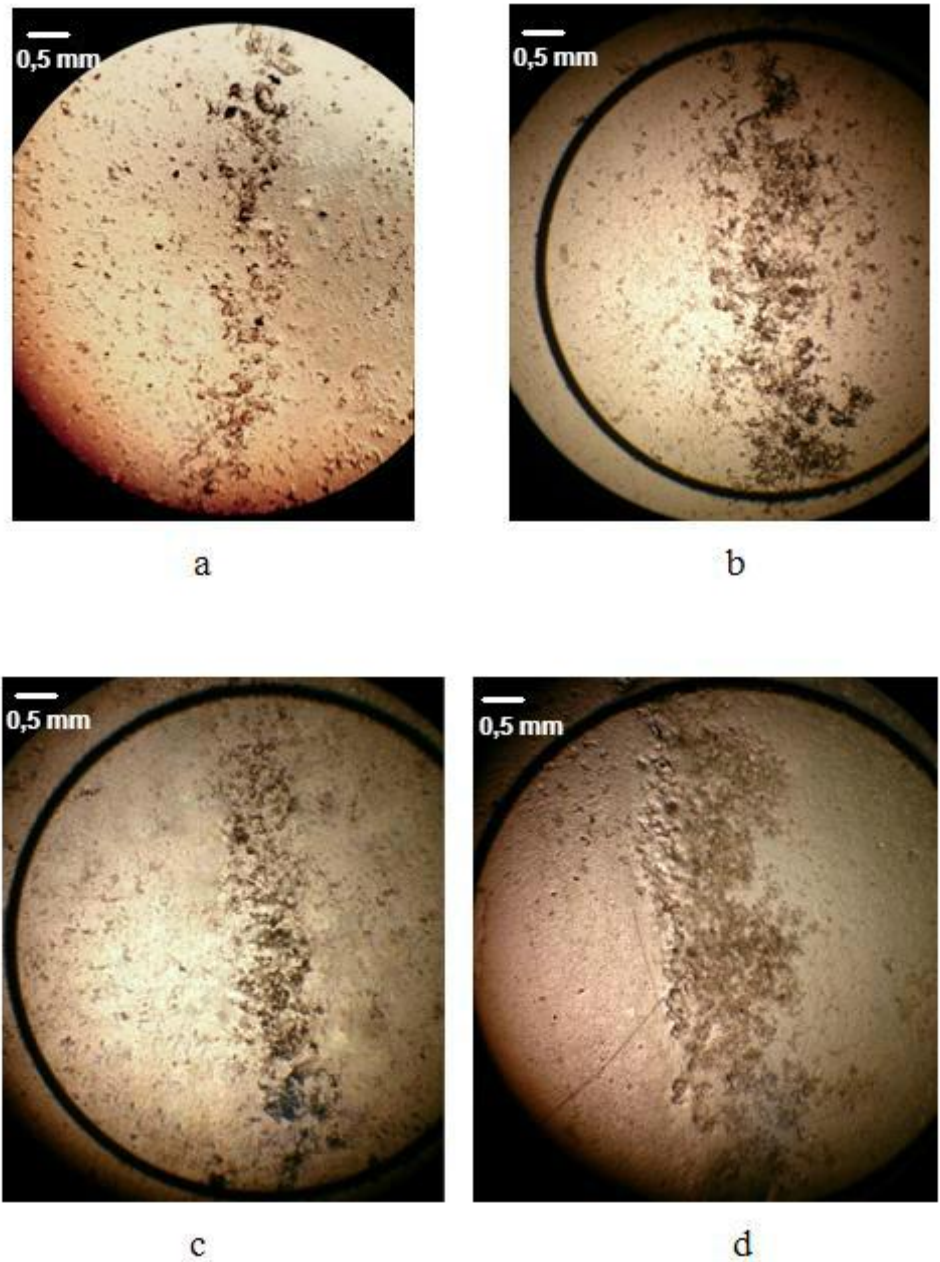

Fig. 4. The results of the study of changes in the deposition of cells of EcN in inhomogeneous MF above the contact surface of the system of two permanent magnets: a) bacteria cells cultivated on standard medium (control); b) bacteria cells cultivated on standard medium with the addition of chelates; c) bacteria cells cultivated on standard medium under the influence of external MF with magnetic field flux density 1500 Oe (0.15 T); d) bacteria cells cultivated on a standard medium with the addition of chelates under the influence of external MF with magnetic field flux density 1500 Oe (0.15 T).

The results were analyzed (Table 2) using the Gwyddion program, which determined the width and surface area of the strips formed by the EcN cell clusters, and the average diameters of the EcN cell clusters. The numbers in brackets in Table 2 show the increase of the respective parameters in comparison with the parameters for bacterial cells cultivated on standard medium (control).

Table 2. The width and surface area of the strips, formed by the EcN cell clusters above the contact surface of the system of two permanent magnets, and the average diameters of the EcN cell clusters.

| The samples cultivated on | The width of the strips, mm | The surface area of the strips, mm² | the average diameters of the EcN cell clusters, micron |
|---|---|---|---|
| a) standard medium (control) | 0,5 | 2.8 | 52,82 |
| b) standard medium with the addition of iron chelate | 1,2 | 7.4 (2,6) | 66,14 (1.25) |
| c) standard medium under the influence of external MF | 0,7 | 3.8 (1,4) | 58,74 (1.1) |
| d) standard medium with the addition of chelates under the influence of external MF | 1,75 | 7.8 (2,8) | 65,74 (1,26) |

The influence of external permanent MF and the addition of iron chelates to the medium of growing the culture of EcN bacteria significantly affects the magnetophoretic mobility and magnetic susceptibility EcN cells (Table. 2). The mean and maximum diameters of EcN cell clusters (Fig. 5, Table 2) cultivated under the influence of an external constant MF are 1.24-1.26 times larger than for the control, and the maximum diameter of the EcN cell clusters differs slightly from the control for samples, cultivated with the addition of iron chelates into the medium (Fig. 5, Table 2).

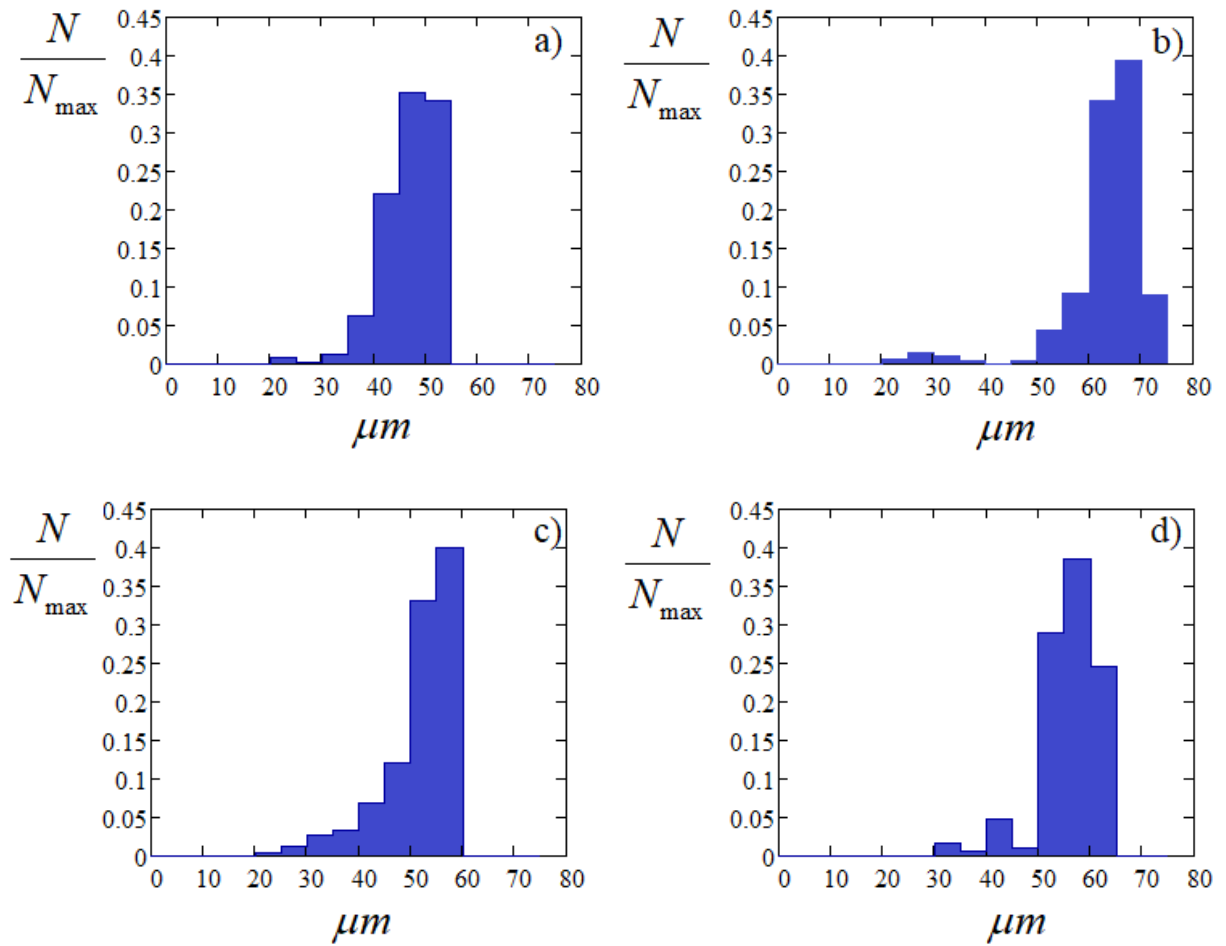

Fig. 5. The distribution of diameters of EcN cell clusters: $N$ is a number of bacterial cell clusters, $N_{max}$ is a total number of bacterial cell clusters analyzed, $D$ is the diameter of bacterial cell cluster; a) - cells of bacteria EcN cultivated on standard medium (control), b) - cells of bacteria EcN cultivated on standard medium with the addition of iron chelate, c) - cells of bacteria EcN cultivated on standard medium under the influence of external permanent MF, d) - cells of bacteria EcN cultivated on standard medium with the addition of iron chelate under the influence of external permanent MF.

The magnetophoretic mobility of four EcN test specimens was compared by measuring the average velocity of bacteria under an inhomogeneous MF at the contact surface of a system of two permanent magnets (Table 3). Table 3 shows the increase in the average velocity of movement of EcN cells relative to the control in parentheses.

The energy $U$ of EcN cell cluster in an external magnetic field can be calculated (using Gaussian) units as

$$U = -\frac{\chi \vec{H}^2}{2} \cdot V_{cl}$$

where $\chi$ is the difference between magnetic susceptibilities of EcN cell cluster and magnetic susceptibility of medium (water), $V_{cl}$ is the volume of EcN cell cluster, $\vec{H}$ is an external magnetic field strength. The volume of spherical EcN cell cluster is $\frac{4\pi r_{cl}^3}{3}$, where $r_{cl}$ is the average radius of EcN cell cluster. The gradient magnetic force acting on EcN cell cluster is

$$\vec{F} = -grad(U).$$

The Stokes force acting on a spherical EcN cell cluster is

$$\vec{F}_{St} = -6\pi\eta r_{cl}\vec{v},$$

where $\eta$ is the dynamic viscosity of liquid medium, the sign "-" means that the Stokes force is antiparallel to the velocity of EcN cell cluster $\vec{v}$. The gradient magnetic force and the Stokes force have the same value for the case of movement of EcN cell cluster with a constant velocity. Then the relation is valid:

$$\vec{v} = \frac{1}{9\eta}\chi\, grad\vec{H}^2 r_{cl}^2$$

The ratio of the values of the magnetic susceptibilities of EcN cells cultivated under different special conditions to the magnetic susceptibility of the cells EcN of the control sample was calculated based on the values of the average velocities of cell movement under gradient magnetic field and taking on account the previous equations:

$$\frac{\chi^{(1)}}{\chi^{(2)}} = \frac{v^{(1)}\left(r_{cl}^{(2)}\right)^2}{v^{(2)}\left(r_{cl}^{(1)}\right)^2},$$

where the indexes $^{(1)}$ and $^{(2)}$ correspond to EcN cells, cultivated under special conditions and standard conditions (control) respectively;

If EcN synthetize single domain BMNs then their magnetization is equal to magnetization saturation of the BMN material $M_0$ and it is oriented parallel to the external magnetic field $\vec{H}$. Than the energy of EcN cell cluster in an external magnetic field can be calculated (using Gaussian) units as

$$U = -\vec{M}_0\vec{H} \cdot V_{BMNs},$$

where $V_{BMNs}$ is the volume BMN material inside EcN cell cluster. The following relation is valid taking on account that the gradient magnetic force is equal the Stokes force

$$\frac{V_{BMNs}^{(1)}}{V_{BMNs}^{(2)}} = \frac{v^{(1)}\left(r_{cl}^{(1)}\right)}{v^{(2)}\left(r_{cl}^{(2)}\right)}.$$

Generalization of the results of the calculation of the average velocities of bacterial cells cultivated under different conditions and the ratios of their magnetic susceptibilities are given in Table. 3.

Table 3. The results of the calculations of the average velocity of movement of cells EcN under the inhomogeneous MF of the system of two permanent magnets and the ratio of their magnetic susceptibilities.

| Cultivation conditions | The average velocity of movement of cells, mm/s | Magnetic susceptibility divided by the magnetic susceptibility of the control |
|---|---|---|
| a) on a standard medium (control) | 0,7 | 1 |
| b) on a standard medium with the addition of iron chelate | 1,6 (2.3) | 1,65 |
| c) in a standard medium under the influence of an external MF | 1 (1.4) | 1,11 |
| d) on a standard medium with the addition of chelates under the influence of an external MF | 1,9 (2.7) | 1,82 |

Table 3 shows that the effect of iron chelate concentration in the medium on the average velocity of cell movement under the inhomogeneous MF exceeds the effect of the external MF on this parameter.

Studies of the influence of the medium with the addition of iron chelate and the external MF on the intensity of cultivation of EcN as a bacterium producing BMN showed that the cultivation of bacteria under the influence of external MF with the 1500 Oe (0.15 T) magnetic field flux density is 14% greater than under control. However, the addition of iron chelate did not have any significant effect on cultivation processes (Fig. 6).

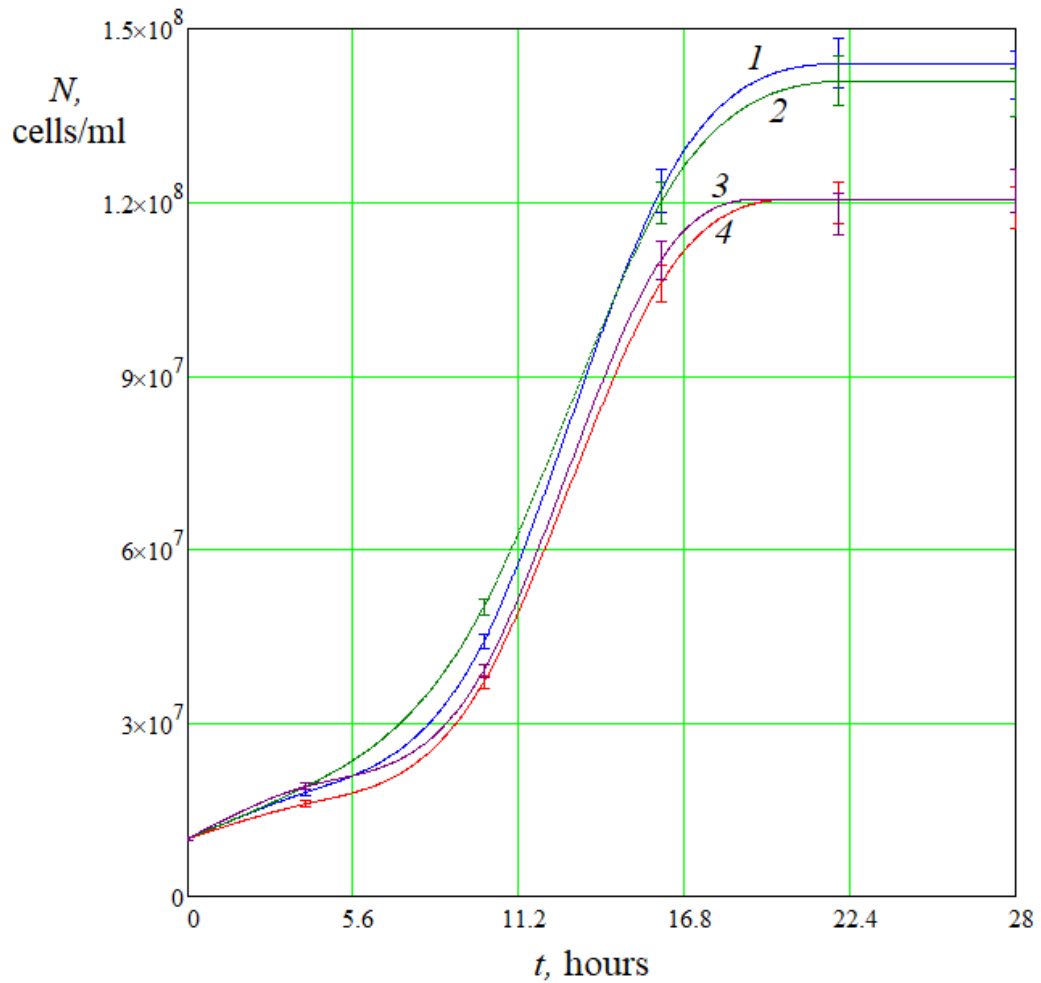

Fig. 6. Growth curves of *E. coli* Nissle 1917 cultivated under different conditions: $N$ is the number of cell clusters per ml of cell suspension, $t$ is the time of cultivation; 1 - cultures cultivated on medium with the addition of iron chelate under the influence of external MF; 2 - cultures cultivated on a standard environment under the influence of an external MF; 3 - cultures cultivated on the medium with the addition of iron chelates; 4 - cultures cultivated on standard environment (control)

Application of an external magnetic field resulted in increase of rate of growth of *E. coli* Nissle 1917. Thus, the study proved the effectiveness of using such conditions of cultivation as the application of an external magnetic field and adding to the nutrient medium of iron chelate to enhance the magnetic susceptibility and magnetophoretic mobility of bacteria producing BMN. As a result, the ways are revealed to create an effective vector for targeting drug delivery of anticancer drugs and therapeutic genes based on the bacterial strain *E. coli* Nissle 1917. The increase of the rate of growth of *E. coli* Nissle 1917 under an external magnetic field can be explained based on the idea about metabolic functions of BMNs that BMNs represent a nano-device for magnetic capture of cluster-type components [35,36]. The rate of growth increases because the size of the zone of capture of cluster-type components increases due to application of a magnetic field [35–37] during cultivation of *E. coli* Nissle 1917.

Nano- or microcontainers are used to immobilize drug molecules on the surface of bacterial cells. Liposomes are most appropriate for this purpose. Liposomes represent microscopically spherical vesicles, usually smaller than 1.5 microns in size, consisting of one or more lipid layers [38]. Unlike other types of nano- and microcontainers (fullerenes, micelles, carbon nanotubes, etc.), liposomes are synthesized from natural phospholipids and are similar in chemical composition to cell membranes and therefore do not require additional modification to acquire biocompatibility, do not cause allergic reactions and are biodegradable. In addition, liposomes are virtually versatile microcontainers because they can carry a wide range of different classes of medical chemotherapy agents. The release of drug in such a targeted delivery system is due to the release of lipase enzymes by cells and liposome-destroying oxidizing agents [39]. Such a mechanism is capable of providing a gradual release of the drug, thereby prolonging its action.

Egg lecithin was used for the synthesis of liposomes. Egg lecithin was isolated as follows. 6 ml of ethyl alcohol was added to 1 g of dry egg yolk. Then the liquid was filtered off and 0.4 ml of acetone was added. The opacification of the solution indicated the deposition of lecithin. The mixture was centrifuged for 3-4 min at 2000 rpm. The supernatant was drained and the precipitated lecithin was dissolved in 10 ml of ethanol and used for liposome synthesis.

Liposomes were obtained by the method of dehydration/rehydration according to the following method: the obtained 1% solution of lecithin in ethyl alcohol was introduced into a round bottom flask of 100 cm$^3$ rotary evaporator and evaporated at a water bath temperature of 60°C to form a lipid film on the walls of the flask. The mass of phospholipids obtained was equal to 2.42 g. Next, 15 ml of centimolar (0.01M) sodium phosphate buffer solution (pH = 7.2-7.4) was added to the flask and shaken for 2 min to form liposomes. The results of liposome formation were verified by light microscopy (Fig. 7).

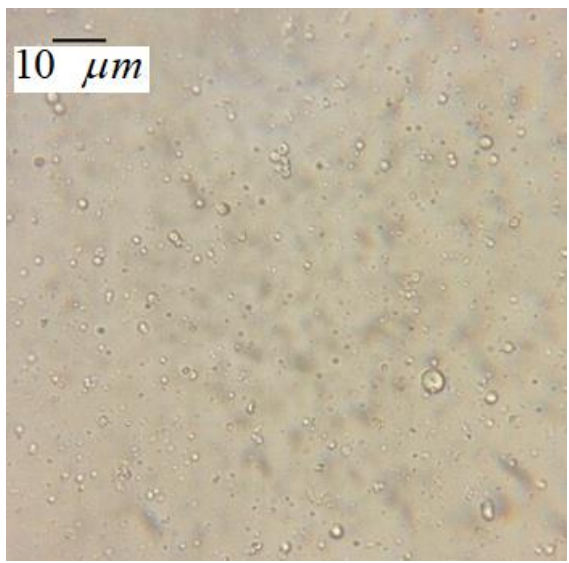

Fig. 7. Micrograph of liposomes.

The size of the liposomes was measured by microwave analysis in Gwyddion. The average value of liposome diameters was equal to about 1 micron.

The surface modification of liposomes was performed with dialdehyde dextran to immobilize liposomes on bacterial cells. The choice of this method to create a targeted delivery system is justified, firstly, by the simplicity of its implementation, and, secondly, by the fact that dialdehydextran is subject to complete biodegradation in the human body unlike other polymers used for liposome surface modification.

The synthesis of dialdehyde dextran by dextran peroxidation was carried out according to the following procedure. 3 ml of 30% hydrogen peroxide solution was added to 15 ml of 5% aqueous dextran solution (40 kDa). The resulting mixture was stirred and poured into a flat bottomed vessel (Petri dishes) so that the thickness of the fluid layer did not exceed 3 mm. The oxidation was carried out in a heat oven at 90°C for 2 hours.

The resulting dry powder of dialdehyde dextran weighing 1.6 g was dissolved in 25 ml of distilled water. 20 ml of this solution was added to 15 ml of the liposomal suspension and stirred with a laboratory stirrer for 1 h at room temperature to form the polymer shell on the liposome surface. Thereafter, another 5 ml of dialdehydextran solution and 12 ml of bacterial suspension at a concentration of $3 \cdot 10^7$ cells/ml were added to the mixture and stirred with a stirrer for 20 min, resulting in the binding of liposomes to bacterial cells.

The formation of liposomal-bacterial complexes was evaluated by light microscopy, the suspension samples were fixed on a slide and stained with fuchsin. In this case, the bacterial cells became dark in color, and the liposomes had a lighter pink color. Micrographs (Fig. 9) show that liposomes bind to bacterial cells, forming complexes.

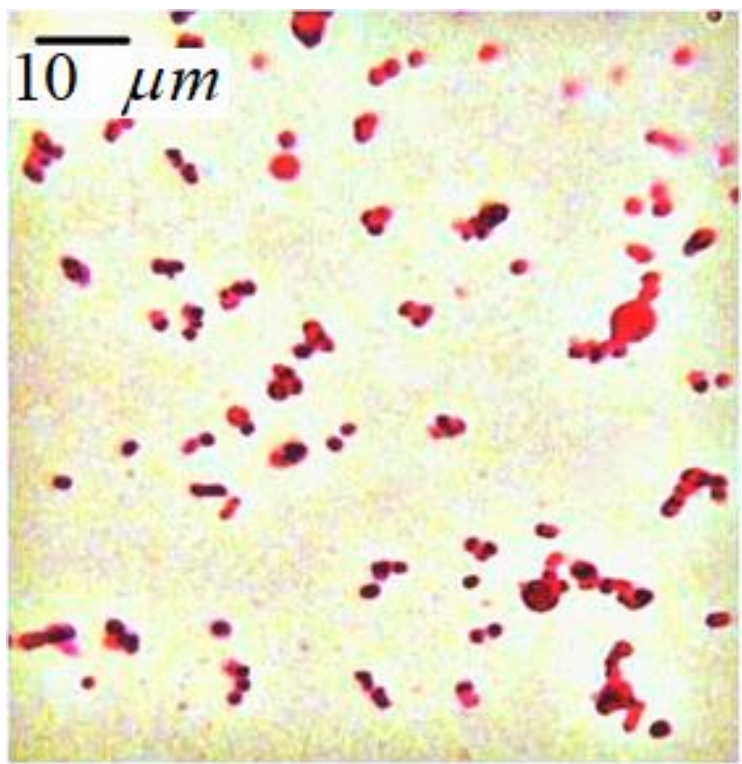

Fig. 9. Micrograph of bacterial-liposomal complexes.

Thus, a targeted drug delivery system was obtained based on the probiotic strain of *E. coli* Nissle 1917 and lecithin liposomes, which can be effectively delivered to the tumor due to magnetophoresis of *E. coli* Nissle 1917 using external control of these complexes by magnetic field.

**Conclusions**

For the first time, it was experimentally discovered that the bacterium *Escherichia coli* Nissle 1917, which is a known vector for drug delivery and gene therapy, has natural magnetically controlled properties (or by other words it is magnetically sensitive), that is, it has sufficient magnetic susceptibility to manifest controlled movement under gradient magnetic field of permanent magnets. Analysis by comparative genomics methods and atomic force and magnetic force microscopy have shown that the natural magnetically controlled properties of *Escherichia coli* Nissle 1917 are based on the fact that it is a BMN producer.

The methods of increase of magnetophoretic mobility and magnetic susceptibility of *Escherichia coli* Nissle 1917 were first discovered in the paper. The quantity of BMNs and magnetic susceptibility of *E. coli* Nissle 1917 cells increase 2-3 times due to cultivation on a nutrient medium with the addition of iron chelate and under the influence of an external magnetic field with a magnetic field flux density of 1500 Oe (0.15 T).

It was also found that the growth rate of *E. coli* Nissle 1917 cells increased by 14% under the influence of an external constant magnetic field of 1500 Oe (0.15 T) and did not change when chelates were added to the culture medium compared to the control.

The results of the paper are important for using *E. coli* Nissle 1917 cell culture for targeted drug and gene delivery using magnetic technology.